\documentclass[12pt]{iopart}

\usepackage{graphicx}
\usepackage{epstopdf}
\usepackage{amssymb}
\usepackage{amsfonts}
\usepackage{color}
\usepackage{ulem}
\usepackage{psfrag}

\begin{document}

\title{Microwave control of Rydberg atom interactions}
\author{S Sevin\c{c}li$^{1,3}$ and T Pohl$^2$}
\address{$^1$Department of Physics and Astronomy, Aarhus University, 8000 Aarhus C, Denmark}
\address{$^2$ Max Planck Institute for the Physics of Complex Systems, N\"othnitzer Strasse 38, 01187 Dresden, Germany}

\ead{sevilaysevincli@iyte.edu.tr}

\begin{abstract}
We investigate the interaction between Rydberg atoms, whose electronic states are dressed by
multiple microwave fields. Numerical calculations are used for an exact description
of the microwave induced interactions, and employed to benchmark a perturbative treatment that yields simple insights into the involved mechanisms. 
Based on this theory, we demonstrate that microwave dressing provides a powerful approach to control dipolar as well as van der Waals interactions 
and even permits to turn them off entirely. In addition, the proposed scheme also opens up possibilities for engineering dominant three-body interactions. 
\end{abstract}


\hspace{15cm}
\address{$3$ Present address: Department of Physics, \.{I}zmir Institute of Technology, IZTECH, TR35430,  \.{I}zmir, Turkey.}
\maketitle

\section{Introduction}
\label{sec:intro}
In recent years the physics of cold Rydberg atoms has attracted widespread interest due to diverse applications, ranging from quantum information science \cite{lfc01,jaksch00,saffman02,dudin12} and quantum simulations of magnetism \cite{weimer10,pode10} over precision measurements \cite{bouchoule02} to nonlinear quantum optics \cite{hofmann13,peyronel12}. Much of these prospects stem from the exaggerated properties of Rydberg states and, in particular, their strong mutual interactions. Due to
the strong $C_6\sim n^{11}$ scaling of the van der Waals coefficient, $C_6$, with the atoms principal quantum 
number, $n$, Rydberg-Rydberg atom interactions can exceed those of ground state atoms by many orders of 
magnitude \cite{gallagher,li05,afrousheh04,mudrich05,carroll04}. Even at comparatively large distances of several $\mu$m the resulting interactions can exceed all remaining energy scales in the system, and, e.g., lead to a strong excitation blockade that inhibits simultaneous excitation of Rydberg atoms within a characteristic blockade radius $R_{\rm bl}$. In addition, the strong $n$-dependence provides a convenient way to tune the interactions and study the transition between the weak and strong interaction regime via moderate changes of the principle quantum number.

On the other hand, the ability to directly tune interactions in a state-selective manner would open up new
possibilities for controlling the properties and time evolution of strongly correlated Rydberg systems and, thereby, add a
valuable degree of flexibility to the aforementioned applications. Here one can make use of the high susceptibility
of Rydberg states to electric and magnetic fields \cite{pss09}, which permits to manipulate their interactions
by applying moderate external fields. Experiments have explored effects of static magnetic and electric \cite{carroll04,bz07,petrus08,han09,park11} fields, where the
latter can be used to induce static dipole moments or dipole-coupled pair resonances, both resulting in direct
dipole-dipole interactions. Time-varying, i.e. microwave fields, offer a more refined way of control, since they
permit to couple different Rydberg levels in a state selective manner. This has been exploited in theoretical studies of polar molecules \cite{buechler07p,buechler07n,gorshkov08}, showing how the combination of a static electric and a microwave field can be used to shape molecular interactions. For cold Rydberg atoms resonant coupling to a single microwave field has been investigated theoretically \cite{muller08} and experimentally \cite{tanas11,moha07,bason08,maxwell13,parades14}, demonstrating an enhanced excitation blockade due to field-induced dipole-dipole interactions. Recent work demonstrates that microwave control of Rydberg states provides a powerful tool, e.g., for manipulating the atomic sensitivity to external fields \cite{jones13}, implementing quantum gates \cite{parades14} or to probe and control Rydberg state dynamics in atomic beam experiments \cite{colombo13}.

In this article, we study far off resonant Rydberg state coupling by a combination of several microwave fields, which is shown to permit versatile control of dipolar as well as van der Waals potentials and to realize conditions with dominating three-body interactions. Numerical calculations
for two interacting atoms combined with a perturbative treatment of many-body systems provide an exact description as
well as an intuitive understanding of the field induced interactions. We discuss several types of realizable interactions and provide simple analytical formulae for the corresponding microwave parameters, including conditions for which the long-distance tail of the interactions can turned off entirely.

The paper is organized as follows. First (section \ref{sec_mw}), we describe the considered system consisting of an ensemble
of Rydberg atoms and introduce the different interaction terms arising from microwave driving and dipole-dipole 
 coupling of different Rydberg states. In section \ref{sec_num} we describe our numerical procedure for calculating the
dressing-induced interaction between two atoms. Finally we present a perturbative many-body treatment
of the resulting interaction, which is compared to the exact calculations and used to identify the required microwave
parameters for obtaining different types of interactions  in section \ref{sec_pert} . 

\begin{figure}[t!]
\begin{center}
\resizebox{0.5\textwidth}{!}{\includegraphics{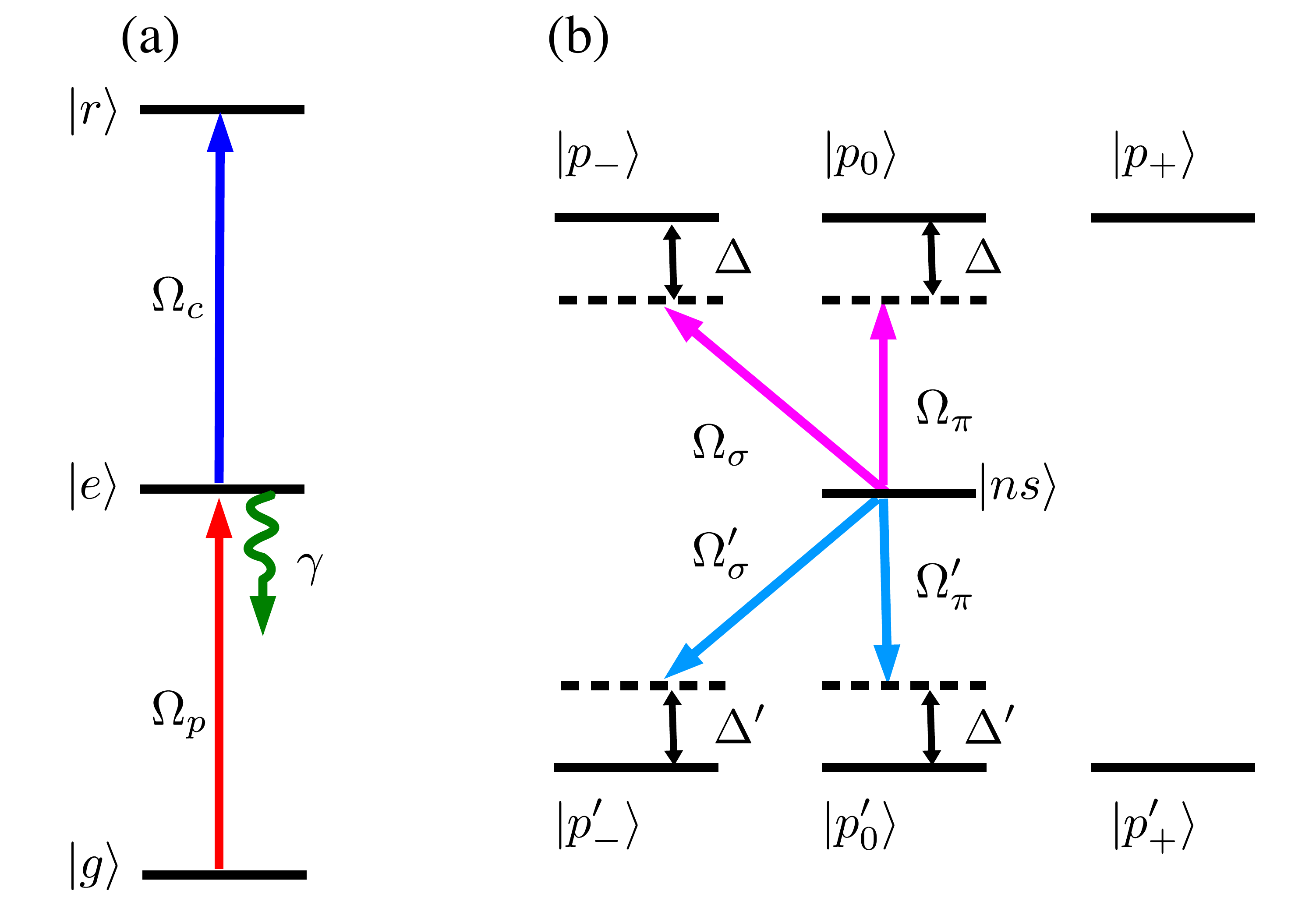}}
\caption{(a)Two-photon excitation scheme for three-level system. States are resonantly coupled by Rabi frequencies $\Omega_p$ and $\Omega_c$ where $\gamma$ stands for spontaneous decay from the intermediate state. (b) Microwave dressing of Rydberg states. Two linearly polarized and two circularly polarized 
fields couple $|ns\rangle$ state to $|p_m\rangle=|npm\rangle$ and $|p^{\prime}_m\rangle=|(n-1)pm\rangle$ 
 states with Rabi frequencies $\Omega_p, \Omega_\sigma$ and $\Omega_p^{\prime}, \Omega_\sigma^{\prime}$; detunings $\Delta$ and $\Delta'$, respectively.  }\label{fig1}
\end{center}
\end{figure}

\section{Microwave dressing of Rydberg states}\label{sec_mw}
We consider an ensemble of cold atoms at positions $\mathbf{R}_i$,
$i = 1, \ldots,N$. We focus on $ns$-Rydberg states ($|r\rangle$) of alkaline atoms, which can be excited from the atomic ground state ($|g\rangle$) by
 two-photon coupling via an intermediate $p$-state ($|e\rangle$) with Rabi frequencies $\Omega_p$ and $\Omega_c$ (see \fref{fig1}a). For alkaline atoms the van der Waals interaction between two $nS_{1/2}$-Rydberg states is repulsive and predominantly originates from
off-resonant dipole-dipole coupling to ($np,(n-1)p$) pair states \cite{saffman10, comparat10}. Four different microwave fields couple the atomic $ns$ state to these two $p$-states. As shown in \fref{fig1}b, we consider two
pairs with linear and circular polarization fields, which have a frequency of $\omega$ and $\omega^\prime$, but different amplitudes
$F_{\pi}$, $F_{\pi}^{\prime}$ and $F_{\sigma}$, $F_{\sigma}^{\prime}$, respectively. The two microwave frequencies for each pair of fields are chosen near-resonant with the transition between the $|ns\rangle$-Rydberg state and the two adjacent $|p\rangle$-states. In particular, $\omega$ is assumed to be close to the $|ns\rangle\rightarrow|np\rangle$  transition energy, while  $\omega^\prime$
is assumed to be close to the $|ns\rangle\rightarrow|(n-1)p\rangle$ transition energy, such that we can define two detunings
$\Delta$ and $\Delta'$ for the upper and lower transitions, respectively. With the radial matrix elements $\mu$ and $\mu'$ of the
$ns \leftrightarrow np$ and $ns \leftrightarrow(n-1)p$ transitions, the field amplitudes define the corresponding the Rabi frequencies  
$\Omega_{\pi}=F_{\pi}\mu/2$, $\Omega_{\pi}^{\prime}=F_{\pi}^{\prime}\mu^{\prime}/2$, $\Omega_{\sigma}=F_{\sigma}\mu/2$ and $\Omega_{\sigma}^{\prime}=F_{\sigma}^{\prime}\mu^{\prime}/2$, as indicated in \fref{fig1}b. In what follows, we will assume the detunings to be larger than the respective Rabi frequencies, i.e. that the atoms are weakly dressed instead of driven resonantly by the microwave fields.

The total Hamiltonian of the system can be split into three parts
\begin{equation}
\hat{H}=\hat{H}_{\rm A}+\hat{H}_{\rm AF}+\hat{H}_{\rm AA}\;.
\end{equation}
The first term denotes the sum of single-atom Hamiltonians, and yields the unperturbed energy spectrum for
each of the $N$ atoms. Following the preceding discussion the atom-microwave coupling is given by
\begin{eqnarray}\label{h_af}
\hat{H}_{\rm AF}=&\sum_i& F_{\pi}z_i\cos(\omega t)+F_{\pi}^{\prime}z_i\cos(\omega^{\prime}t)+F_{\sigma}\left[x_i\cos(\omega t)+y_i\sin(\omega t)\right]\nonumber\\
&+&F_{\sigma}^{\prime}\left[x_i\cos(\omega^{\prime}t)+y_i\sin(\omega^{\prime}t)\right]\;,
\end{eqnarray}
where the field amplitudes and frequencies have been defined above. Finally, the dipole-dipole interaction between the
Rydberg atoms is given by
\begin{equation}\label{h_int}
\hat{H}_{\rm AA}=\sum_{i<j}\frac{{\bf r}_i{\bf r}_j}{R_{ij}^3}-3\frac{({\bf r}_i{\bf R}_{ij})({\bf r}_j{\bf R}_{ij})}{R_{ij}^5}\:,
\end{equation}
where ${\bf R}_{ij}={\bf R}_{j}-{\bf R}_{i}$ denotes the distance vector between two Rydberg atoms at positions ${\bf R}_i$ and ${\bf R}_j$.

For a description of $ns-ns$ interactions we may approximately neglect spin-orbit coupling, e.g. the fine structure of the Rydberg states \cite{zimmerman79,li03}, such that each atomic state is characterized by the three quantum numbers, n, l and m. Denoting the relevant Rydberg states by  $|s\rangle\equiv|n,l=0,m=0\rangle$, $|p_{0,\pm}\rangle\equiv|n,l=1,m=0,\pm 1\rangle$ and $|p'_{0,\pm}\rangle\equiv|n-1,l=1,m=0,\pm 1\rangle$, we can write the
isolated-atom Hamiltonian in our reduced basis as
\begin{equation}
\hat{H}_{\rm A}=E_{p}\sum_{i}(\hat{\sigma}_{p_-p_-}^{(i)}+\hat{\sigma}_{p_0p_0}^{(i)}+\hat{\sigma}_{p_+p_+}^{(i)})+E_{p^{\prime}}\sum_{i}(\hat{\sigma}_{p_-^{\prime}p_-^{\prime}}^{(i)}+\hat{\sigma}_{p_0^{\prime}p_0^{\prime}}^{(i)}+\hat{\sigma}_{p_+^{\prime}p_+^{\prime}}^{(i)})\:,
\end{equation}
where $\hat{\sigma}_{\alpha\alpha}^{(i)}=|\alpha^{(i)}\rangle\langle\alpha^{(i)}|$ denote single-atom projection
operators of the $i$th atom onto a given state $|\alpha\rangle=|l,m,n\rangle$. The energies  $E_{p}$ and $E_{p}^{\prime}$ are the energies of the $np$-states and $(n-1)p$ states relative to the $ns$-Rydberg state, respectively, and define the microwave
detunings $\Delta=\omega-|E_{p}|$ and $\Delta^{\prime}=\omega^{\prime}-|E_{p^{\prime}}|$.

\section{Numerical description}\label{sec_num}
In addition to an analytical perturbative treatment of the microwave-induced interactions, we also performed numerical calculations that account for all possible Rydberg state couplings due to microwave driving and dipole interactions. To this end, we use a Floquet representation \cite{chu04} of the Hamiltonian, which permits to eliminate the time-dependence in  (\ref{h_af}) and yields a time-independent Floquet matrix eigenvalue problem. For two atoms the corresponding Hilbert space is spanned by the basis states $|\alpha^{(1)},\alpha^{(2)},\nu_\pi,\nu_\sigma,\nu_\pi^\prime,\nu_\sigma^\prime\rangle$ where $\alpha^{(i)}$ refers to the atomic states of each of the two atoms and the quantum numbers $\nu_\pi$, $\nu_\pi^\prime$, $\nu_\sigma$ and $\nu_\sigma^\prime$ correspond to the number of microwave photons with the respective polarization and frequency. Since the chosen Rabi frequencies are  considerably smaller than the respective transition energies it suffices to consider three adjacent photon numbers $\nu=-1,0,1$ for each microwave mode, which makes for a
total of $3969$ pair states. Upon diagonalizing the resulting $3969\times3969$ Floquet matrix, we obtain the microwave-dressed interaction potentials and corresponding molecular eigenstates, among which we are interested in the ($ns-ns$)-asymptote, i.e. the molecular potential curves that asymptotically converges to $|s,s,0,0,0,0\rangle$ upon decreasing the microwave amplitudes.

\section{Perturbative approach}\label{sec_pert}
In order to develop an analytical description we first transform into a frame of reference
 that rotates with the applied fields at the frequencies $\omega$ and $\omega'$, by transforming the $N$-atom
Hamiltonian according to $\hat{H}\rightarrow\hat{U}\hat{H}\hat{U}^{\dagger}+i\frac{\partial \hat{U}}{\partial t}\hat{U}^{\dagger}$ where
\begin{equation}
U=\exp\left[\sum_i i\omega t(\hat{\sigma}_{p_-p_-}^{(i)}+\hat{\sigma}_{p_0p_0}^{(i)}+\hat{\sigma}_{p_+p_+}^{(i)})-i\omega^{\prime} t(\hat{\sigma}_{p_-^{\prime}p_-^{\prime}}^{(i)}+\hat{\sigma}_{p_0^{\prime}p_0^{\prime}}^{(i)}+\hat{\sigma}_{p_+^{\prime}p_+^{\prime}}^{(i)})\right]\:.
\end{equation}
The transformed Hamiltonian contains time dependent terms that vary proportional to $e^{2i\omega t}$, $e^{2i\omega^{\prime} t}$ and $e^{\pm i(\omega-\omega^{\prime}) t}$. If the difference between  $\omega$ and $\omega'$ is sufficiently large we can neglect all of these fast oscillating terms to obtain a time-independent Hamiltonian within the rotating wave approximation. The single atom contribution is of the familiar form
\begin{equation}
\hat{H}_A=-\Delta\sum_{i}(\hat{\sigma}_{p_-p_-}^{(i)}+\hat{\sigma}_{p_0p_0}^{(i)}+\hat{\sigma}_{p_+p_+}^{(i)})+\Delta^{\prime}\sum_{i}(\hat{\sigma}_{p_-^{\prime}p_-^{\prime}}^{(i)}+\hat{\sigma}_{p_0^{\prime}p_0^{\prime}}^{(i)}+\hat{\sigma}_{p_+^{\prime}p_+^{\prime}}^{(i)})\:,
\end{equation}
and the atom field coupling is given by
\begin{eqnarray}
\hat{H}_{AF}&=&\Omega_{\pi}\sum_{i}(\hat{\sigma}_{sp_0}^{(i)}+\hat{\sigma}_{p_0s}^{(i)})+\sqrt{2}\Omega_{\sigma}\sum_{i}(\hat{\sigma}_{sp_+}^{(i)}+\hat{\sigma}_{p_+s}^{(i)})\nonumber\\
&+&\Omega_{\pi}^{\prime}\sum_{i}(\hat{\sigma}_{sp_0^{\prime}}^{(i)}+\hat{\sigma}_{p_0^{\prime}s}^{(i)})+\sqrt{2}\Omega_{\sigma}^{\prime}\sum_{i}(\hat{\sigma}_{sp_+^{\prime}}^{(i)}+\hat{\sigma}_{p_+^{\prime}s}^{(i)})\:.
\end{eqnarray}
The resulting expression for the Rydberg-Rydberg atom interaction Hamiltonian (\ref{h_int}) is given in \ref{apa}.

\begin{figure}[t!]
\begin{center}
\resizebox{0.5\textwidth}{!}{\includegraphics{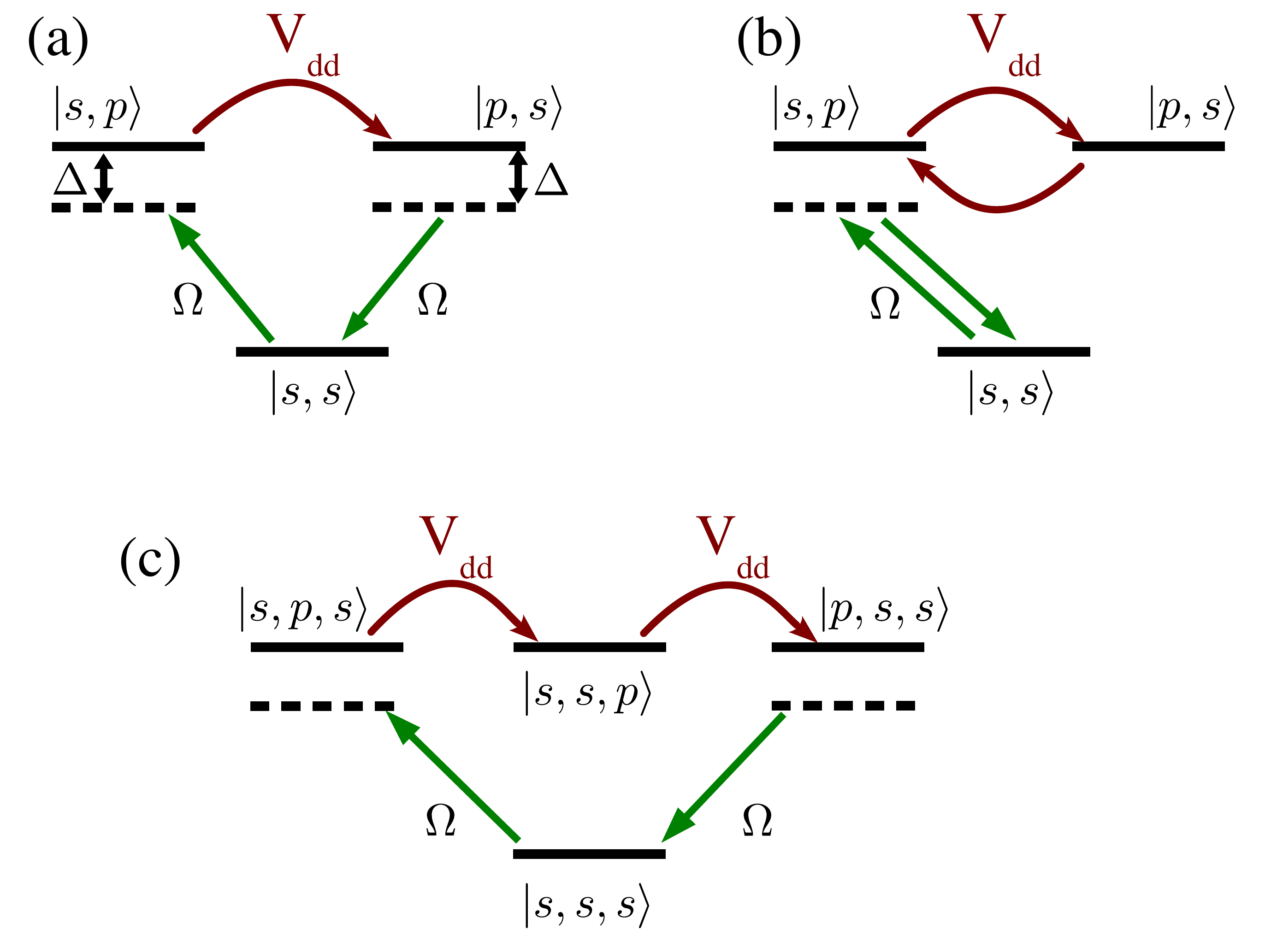}}
\caption{Schematic explanation of various exchange mechanisms between the pairs, which is controlled by microwave fields, result in different types of interactions. A single excitation exchange between the pairs gives rise to  (a) dipolar interactions which is manifested in the third-order perturbation theory. Similarly, two exchange processes yield (b) van der Waals and (c) three-body interactions in the fourth order contribution between two and three atoms, respectively. }\label{fig2}
\end{center}
\end{figure}

In the limit where both the Rabi frequencies as well as the dipole-dipole couplings are smaller than the microwave detunings we can apply perturbation theory to calculate the long-range part of the dressing-induced interactions. The first order contribution vanishes. 
In the laboratory frame, the second order energy correction yields the standard van der Waals interaction \cite{singer05,walker08}
\begin{equation}\label{eq:int_lab}
E_{\rm lab}^{(2)}=\sum_{i< j}\frac{C_6}{R_{ij}^6}\;,
\end{equation}
with $C_6=-12\mu^2\mu^{\prime2}/(E_p+E_{p^{\prime}})$. In the rotating frame of reference, the second order energy correction
\begin{equation}
E^{(2)}=N\left(\frac{\Omega_{\pi}^2}{\Delta}-\frac{\Omega_{\pi}^{\prime2}}{\Delta^{\prime}}+\frac{2\Omega_{\sigma}^2}{\Delta}-\frac{2\Omega_{\sigma}^{\prime2}}{\Delta^{\prime}}\right)
\end{equation}
just yields the combined light shift of the applied microwave fields. Atomic interactions arise from the microwave dressing in third order
\begin{eqnarray}\label{E3}
\fl E^{(3)}&=&\sum_{i<j}\frac{2\mu^2}{R_{ij}^3}\left[\left(\frac{\Omega_{\pi}}{\Delta}\right)^2\left(1-3\frac{Z_{ij}^2}{R_{ij}^2}\right)+\left(\frac{\Omega_{\sigma}}{\Delta}\right)^2\left(2-3\frac{X_{ij}^2+Y_{ij}^2}{R_{ij}^2}\right)-6\frac{\Omega_{\pi}\Omega_{\sigma}}{\Delta^2}\frac{X_{ij}Z_{ij}}{R_{ij}^2}\right]\nonumber\\
\fl &&+\sum_{i<j}\frac{2\mu^{\prime2}}{R_{ij}^3}\left[\left(\frac{\Omega_{\pi}^{\prime}}{\Delta^{\prime}}\right)^2\left(1-3\frac{Z_{ij}^2}{R_{ij}^2}\right)+\left(\frac{\Omega_{\sigma}^{\prime}}{\Delta^{\prime}}\right)^2\left(2-3\frac{X_{ij}^2+Y_{ij}^2}{R_{ij}^2}\right)-6\frac{\Omega_{\pi}^{\prime}\Omega_{\sigma}^{\prime}}{\Delta^{\prime2}}\frac{X_{ij}Z_{ij}}{R_{ij}^2}\right]\:,
\end{eqnarray}
and fourth order
\begin{eqnarray}\label{E4}
\fl E^{(4)}&=&\sum_{i<j}\frac{2\mu^4}{\Delta R_{ij}^6}\left[\left(\frac{\Omega_{\pi}}{\Delta}\right)^2\left(1+3\frac{Z_{ij}^2}{R_{ij}^2}\right)+\left(\frac{\Omega_{\sigma}}{\Delta}\right)^2\left(2+3\frac{X_{ij}^2+Y_{ij}^2}{R_{ij}^2}\right)+6\frac{\Omega_{\pi}\Omega_{\sigma}}{\Delta^2}\frac{X_{ij}Z_{ij}}{R_{ij}^2}\right]\nonumber\\
\fl &&-\sum_{i<j}\frac{2\mu^{\prime4}}{\Delta^{\prime} R_{ij}^6}\left[\left(\frac{\Omega_{\pi}^{\prime}}{\Delta^{\prime}}\right)^2\left(1+3\frac{Z_{ij}^2}{R_{ij}^2}\right)+\left(\frac{\Omega_{\sigma}^{\prime}}{\Delta^{\prime}}\right)^2\left(2+3\frac{X_{ij}^2+Y_{ij}^2}{R_{ij}^2}\right)+6\frac{\Omega_{\pi}^{\prime}\Omega_{\sigma}^{\prime}}{\Delta^{\prime2}}\frac{X_{ij}Z_{ij}}{R_{ij}^2}\right]\nonumber\\
\fl &+&\sum_{i\neq j,k}^N U_{3b}(\mathbf{R}_i,\mathbf{R}_j,\mathbf{R}_k)\:,
\end{eqnarray}
perturbation theory.
The last term in (\ref{E4}) corresponds to microwave-induced three-body interactions (see \ref{apb}), that can not
be written in terms of binary potentials. The interactions of the third order contribution have dipolar character, which arise from a single excitation exchange between ($s, p$)-pairs that are virtually populated by the off-resonant microwave fields (see \fref{fig2}a). As illustrated in \fref{fig2}b the next order binary terms correspond to two consecutive exchange processes between an atom pair and therefore results in a van der Waals type interaction $\sim R_{ij}^{6}$ (see \ref{E4}). In a similar fashion, the three-body terms also involve two pair-exchange processes but between three atoms, as illustrated in \fref{fig2}c. Specific choices of the
microwave parameters significantly simplify the derived expressions (\ref{E3}) and (\ref{E4}), and permit to realize qualitatively different types of interactions, which will be discussed in the following.

\subsection{Dipole-dipole interaction}

We start by considering only the two linearly polarized fields, i.e. $\Omega_{\sigma}=\Omega_{\sigma}^{\prime}=0$. In this most simple case, the interactions (\ref{E3}) and (\ref{E4}) greatly simplify and depend on the molecular orientation only through the $Z$-component of the inter-particle distance. If we further require for the microwave parameters
\begin{equation}\label{condd}
 \Omega^{\prime}=\frac{|\mu|\Delta^{\prime}\Omega}{|\mu^{\prime}|\Delta}\:\:,\quad \Delta^{\prime}=\frac{\mu^{\prime2}\Delta}{\mu^2},
\end{equation}
the fourth order contribution vanishes, $E^{(4)}=0$, and the third order term yields pure dipole-dipole interactions
\begin{equation}\label{e_dd}
 E^{(3)}=\sum_{i<j}\frac{\Omega^2}{\Delta^2}\frac{4\mu^4}{R_{ij}^3}\left(1-3\frac{Z_{ij}^2}{R_{ij}^2}\right)\:.
\end{equation}
The only remaining free parameter that controls the strength of the microwave induced
interactions is the ratio of the Rabi frequency and the detuning, $\Omega/\Delta$.

\begin{figure}[t!]
\begin{center}
\resizebox{0.5\textwidth}{!}{\includegraphics{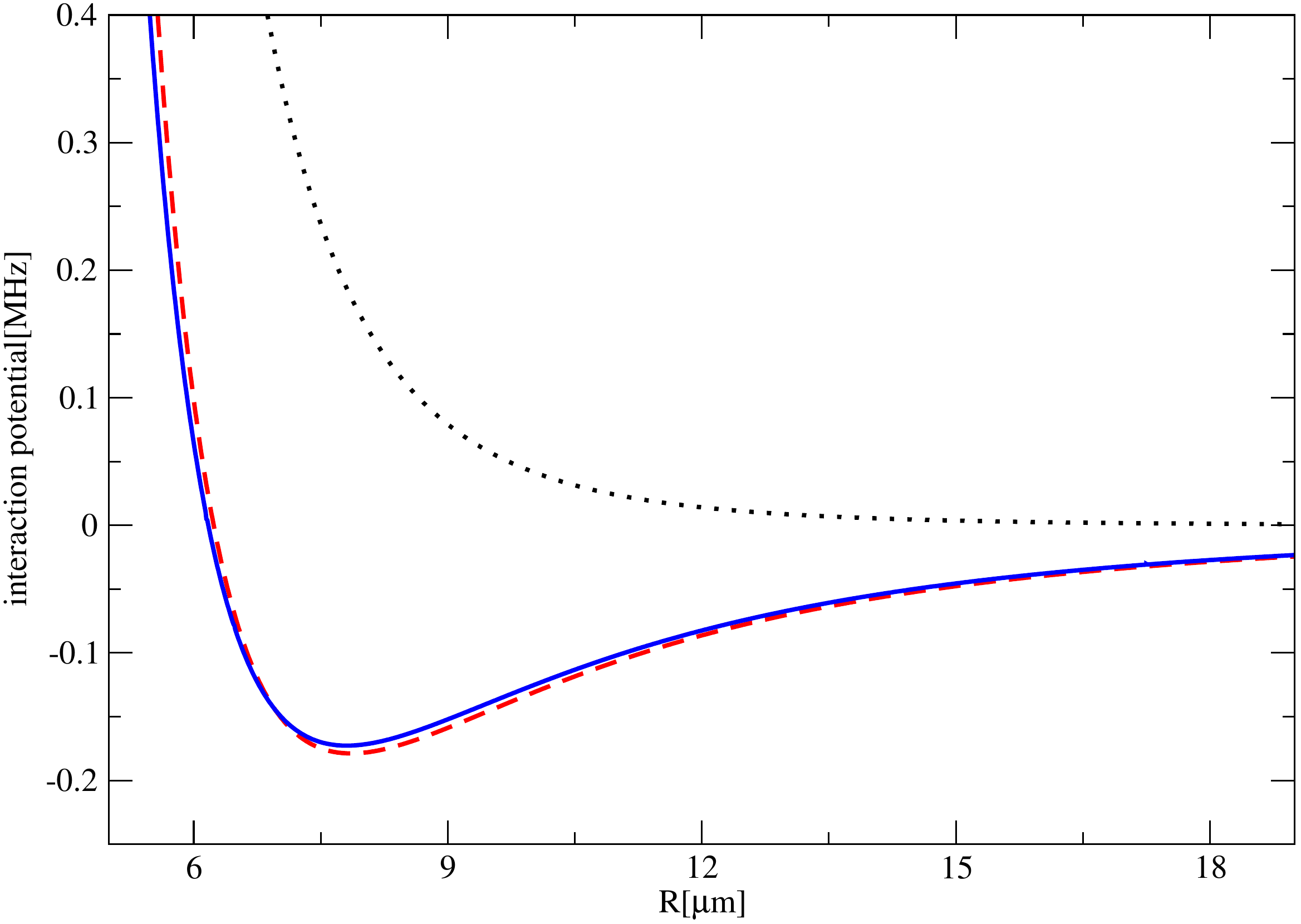}}
\caption{Perturbative potential (dashed red line) compared to exact Floquet calculations (blue line) where the assumptions in (\ref{condd})
are made for $n=55$ and $|\Omega/\Delta|=0.08$ and $\Delta=-150$ MHz. The dotted black line shows the bare vdW interaction.
 }\label{fig3}
\end{center}
\end{figure}

To obtain the total atomic interaction we add both contributions in the laboratory frame (\ref{eq:int_lab}) and rotating frame of reference (\ref{e_dd}). \Fref{fig3} shows the resulting interaction potential between two Rubidium atoms in $55s$ Rydberg states, microwave dressed according to (\ref{condd}) for $|\Omega/\Delta|=0.08$ and $\Delta=-150$ MHz. A comparison to the numerical result demonstrates that this procedure indeed yields a proper description of the total interactions. Here we assumed that the internuclear axis is aligned with the polarization axis of the microwave, such that (\ref{e_dd}) is attractive. A crossover between van der Waals interactions and dipole-dipole interactions due to the microwave appears at a critical distance $\sim(C_6\Delta^2/(8\mu^4\Omega^2))^{1/3}$. Hence, the combination of the attractive dipole-dipole interaction with the repulsive van der Waals term gives rise to a pronounced potential well around this distance, whose depth and position can be widely tuned by the microwave parameters. Since the total potential turns isotropically repulsive for perpendicular orientation of the internuclear axis this type of potential can be used to create long-range bound molecules \cite{bois02,gallmol} which can be aligned with the polarization direction and whose properties can be well controlled by the external microwave fields and tuned dynamically. Similar interaction potentials have also been considered for microwave-dressed polar molecules \cite{buechler07n,cooper09} and where shown to enable superfluid p-wave pairing in ultracold fermionic ensembles \cite{cooper09}.
 
\subsection{van der Waals interaction}

\begin{figure}[t!]
\begin{center}
\resizebox{0.5\textwidth}{!}{\includegraphics{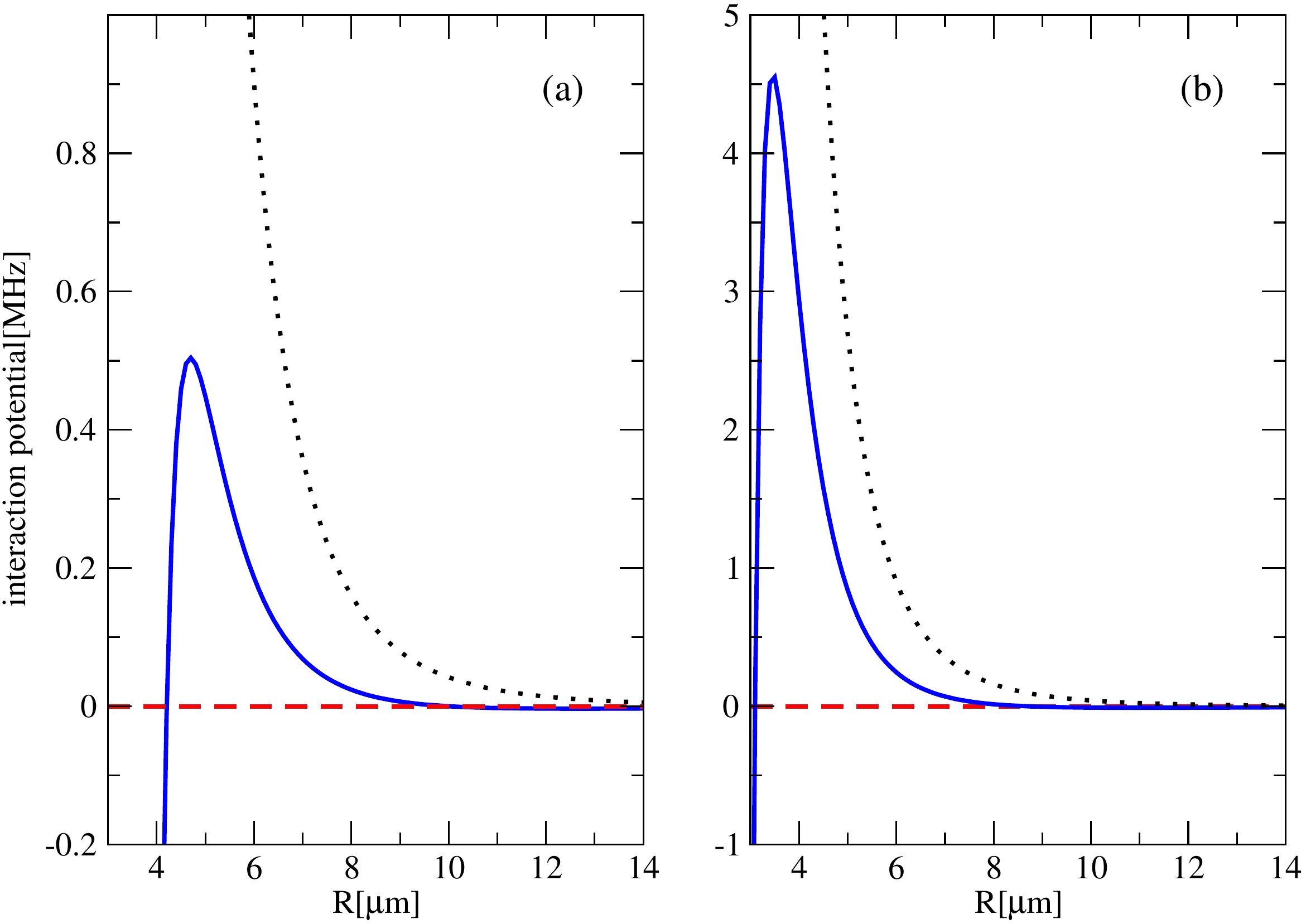}}
\caption{Perturbative potential (dashed red line) compared to the bare vdW interaction 
(dotted black line) and the nonperturbative Floquet calculations (blue line). We have chosen 
$n=55$ and $|\Omega/\Delta|=0.15$ (a), $0.2$ (b) for $\Delta=-150$MHz, while the remaining parameters are chosen according to  (\ref{convdw})
to cancel the long-range interaction.
 }\label{fig4}
\end{center}
\end{figure}

Another important case can be realized by applying both linearly and circularly polarized fields. In order to
simplify the discussion we will set the Rabi frequencies to
\begin{eqnarray}\label{con_om}
\Omega_{\sigma}&=&\Omega_{\pi}=\Omega\nonumber\\
\Omega_{\sigma}^\prime&=&\Omega_{\pi}^{\prime}=\Omega^{\prime}\:.
\end{eqnarray}
If one further requires a slightly different relation between the parameters for the driving fields of the upper and lower transition
\begin{equation}\label{convdw}
 \Omega^{\prime}=\frac{|\mu|\Delta^{\prime}\Omega}{|\mu^{\prime}|\Delta}\:\:,\quad \Delta^{\prime}=-\frac{\mu^{\prime2}\Delta}{\mu^2},
\end{equation}
one can cancel the third order contribution $E^{(3)}=0$, i.e. realize vanishing dipole-dipole interactions in the presence of microwave driving. However, (\ref{E4}) remains finite and yields isotropic van der Waals type interactions. Hence the total interaction in the rotating and laboratory frame can be expressed in terms of an effective van der Waals interaction
\begin{equation}\label{E_vdw}
E=\sum_{i<j}\frac{C_6^{\prime}}{R_{ij}^6}\:,
\end{equation}
with a van der Waals coefficient
\begin{equation}\label{c_eff}
 C_6^{\prime}=C_6+24\kappa\frac{\mu^4}{\Delta}\:,
\end{equation}
that can be tuned by the external fields. Again the dressing-induced contribution is suppressed by the ratio $\kappa=\Omega^2/\Delta^2$, which corresponds to the small fraction of admixed  $p$-states. However, the microwave detuning, $\Delta$, which enters in the denominator of (\ref{c_eff}), can be made much smaller than the energy mismatch between the different pair states in the laboratory frame. The latter determines the strength of the bare van der Waals interaction, such the magnitude of the microwave-induced van der Waals coefficient can be on the same order than the bare coefficient. However, the additional potential can be made attractive or repulsive depending on the sign of the microwave detuning. In particular, one can, thus, use the microwave fields to cancel the binary van der Waals interactions entirely by setting
\begin{equation}\label{cond_van}
\Delta=-24\kappa\mu^4/C_6. 
\end{equation}
\Fref{fig4} shows the resulting interaction potentials for microwave dressed Rb($55s$) atoms with $|\Omega/\Delta|=0.15, 0.2$ and $\Delta=-150$MHz. A comparison to the numerical results and the bare van der Waals interaction demonstrates that the long-range tail of the latter can indeed be suppressed to vanishingly small values. However, below a critical distance the dipole-dipole coupling between the Rydberg states exceeds the microwave detuning, causing a break down
of the perturbation theory. As result, the interaction starts to increase again for distances within this critical radius (see Fig. \ref{fig4}b).

\begin{figure}[t!]
\begin{center}
\resizebox{0.5\textwidth}{!}{\includegraphics{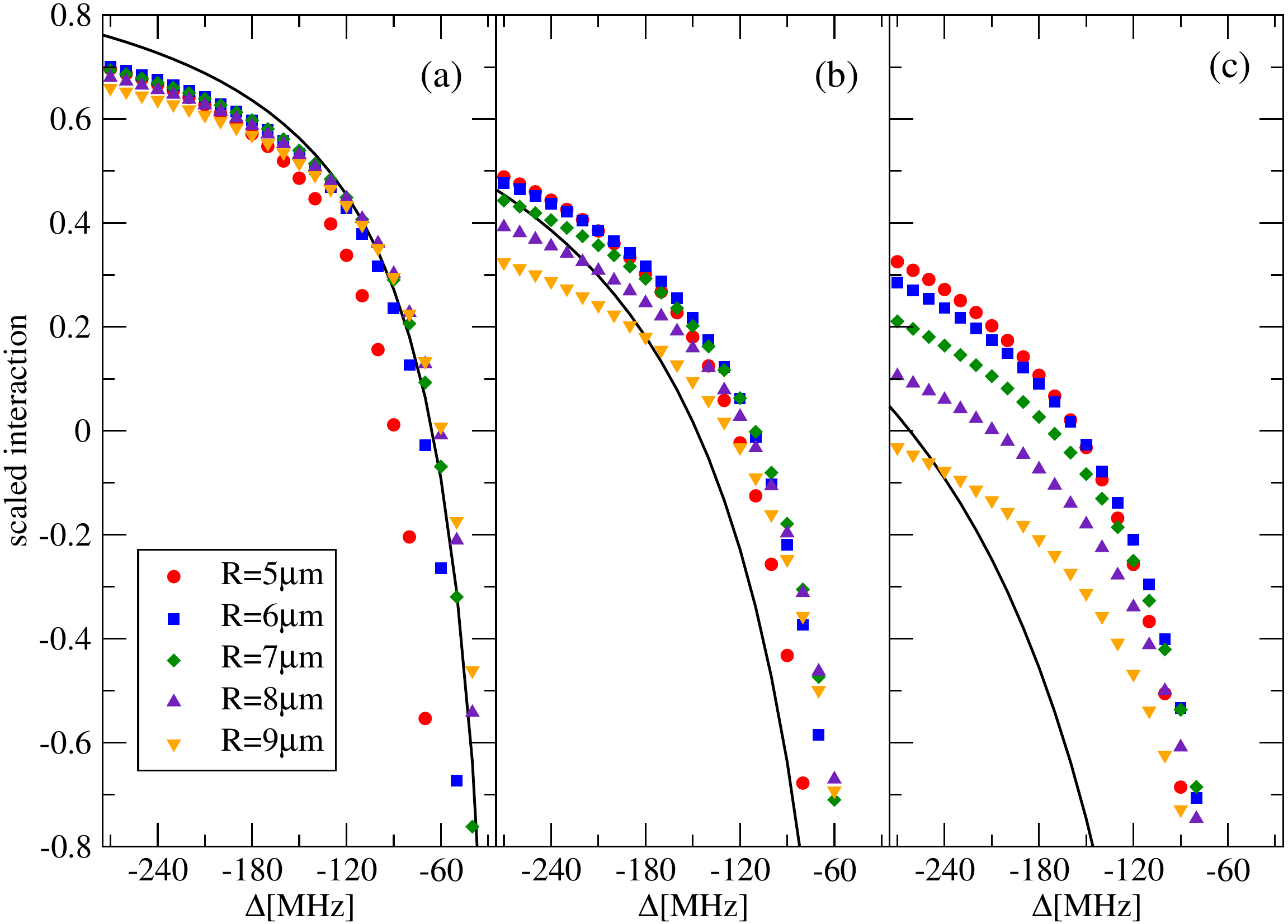}}
\caption{Interaction energies relative to bare vdW interaction $V=C_6/R^6$ at different interatomic distances and 
for different ratios of $|\Omega/\Delta|=0.1$ (a), $0.15$ (b), $0.2$ (c)  and for $n=55$. The solid line shows 
the perturbative result. }\label{fig5}
\end{center}
\end{figure}

\begin{figure}[h!]
\begin{center}
\resizebox{0.5\textwidth}{!}{\includegraphics{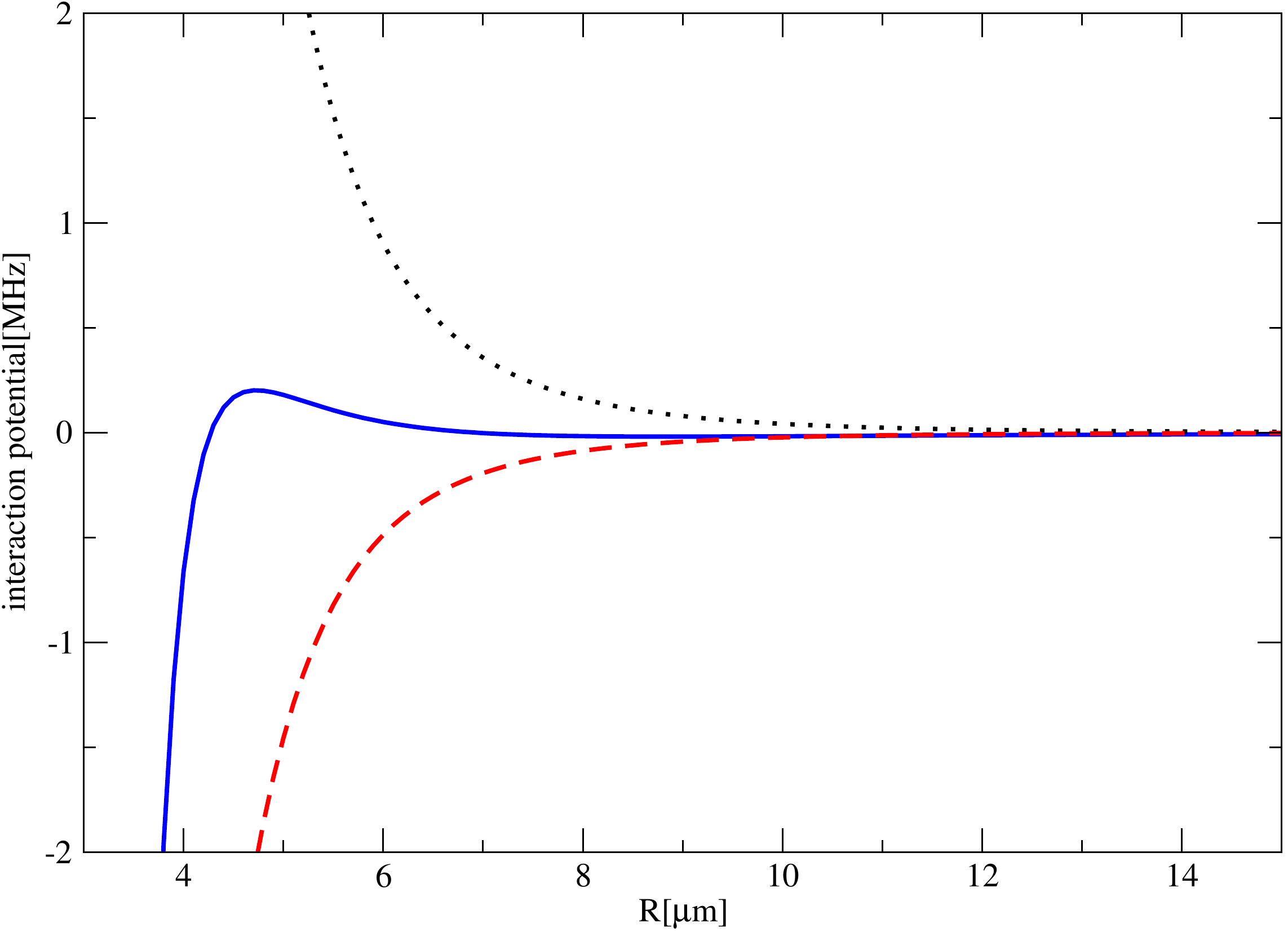}}
\caption{Perturbative potential (dashed red line) compared to the bare vdW interaction 
(dotted black line) and the nonperturbative Floquet calculations (blue line) for optimized detuning $\Delta=-170$ MHz, $|\Omega/\Delta|=0.2$ and for $n=55$.  }\label{fig6}
\end{center}
\end{figure}

To further analyze the possibility to cancel interactions, we show in Fig.\ref{fig5} the $\Delta$-dependence of the numerically obtained interaction energies for different interatomic distances relative to the bare interaction energy (\ref{eq:int_lab}). With a decreasing ratio $\Omega/\Delta$ the data points tend to collapse onto a single curve that approaches the analytical prediction (\ref{E_vdw}) and (\ref{c_eff}) for large distances. Despite showing significant deviations from the perturbative results, comparatively large ratios still yield detunings for which the interactions are strongly suppressed down to comparably small distances of a few micrometers. For given microwave intensities one can, hence, find an optimal detuning for which interactions are strongly suppressed over a broad range of distances. One such case is shown in \fref{fig6}, corresponding to  $|\Omega/\Delta|=0.2$ and $\Delta=-170$ MHz.

The found modification of the interaction potential can have significant consequences for the Rydberg blockade effect. For example, for Rydberg excitation with a typical Rabi frequency $\sim200$kHz and the interaction curves of \fref{fig6}, microwave dressing decreases the corresponding blockade radius by about a factor two. This implies a significant change of the excitation dynamics, which for the bare van der Waals interaction would require a $\sim4000$-fold increase of the intensity of the laser that drives the Rydberg transition. Such a reduction of the blockade radius implies a eightfold decrease of the fraction of blockaded atoms, a quantity which has been thoroughly studied in a number of previous measurements and, thereby, provides a sensitive and well established experimental approach to study the effects of the microwave dressing and to identify optimal field parameters for maximum interaction control.

\subsection{Three-body interaction}
Three-body interactions that appear in the fourth order energy correction  (see (\ref{E4})) become significant  under conditions where binary interactions are minimized. The general  expression for $U_{\rm 3b}$ is rather lengthy and given in \ref{apb}. However, if we choose the parameters such that the long-range tail of the binary interactions vanishes ( conditions (\ref{con_om}), (\ref{convdw}) and (\ref{cond_van})) the potential takes on a particularly simple form

\begin{equation}
\fl U_{\rm 3b}(\mathbf{R}_i,\mathbf{R}_j,\mathbf{R}_k)=\sum_{i<j<k}-\frac{6\Omega^2}{\Delta^3}\mu^4\left\{\frac{1}{R_{ij}^3R_{jk}^3}\left[1-3\frac{(\mathbf{R}_{ij}\cdot\mathbf{R}_{jk})^2}{R_{ij}^2R_{jk}^2}\right]
+i\leftrightarrow j+j\leftrightarrow k\right\}\:.
\end{equation}

\begin{figure}[t!]
\begin{center}
\resizebox{0.6\textwidth}{!}{\includegraphics{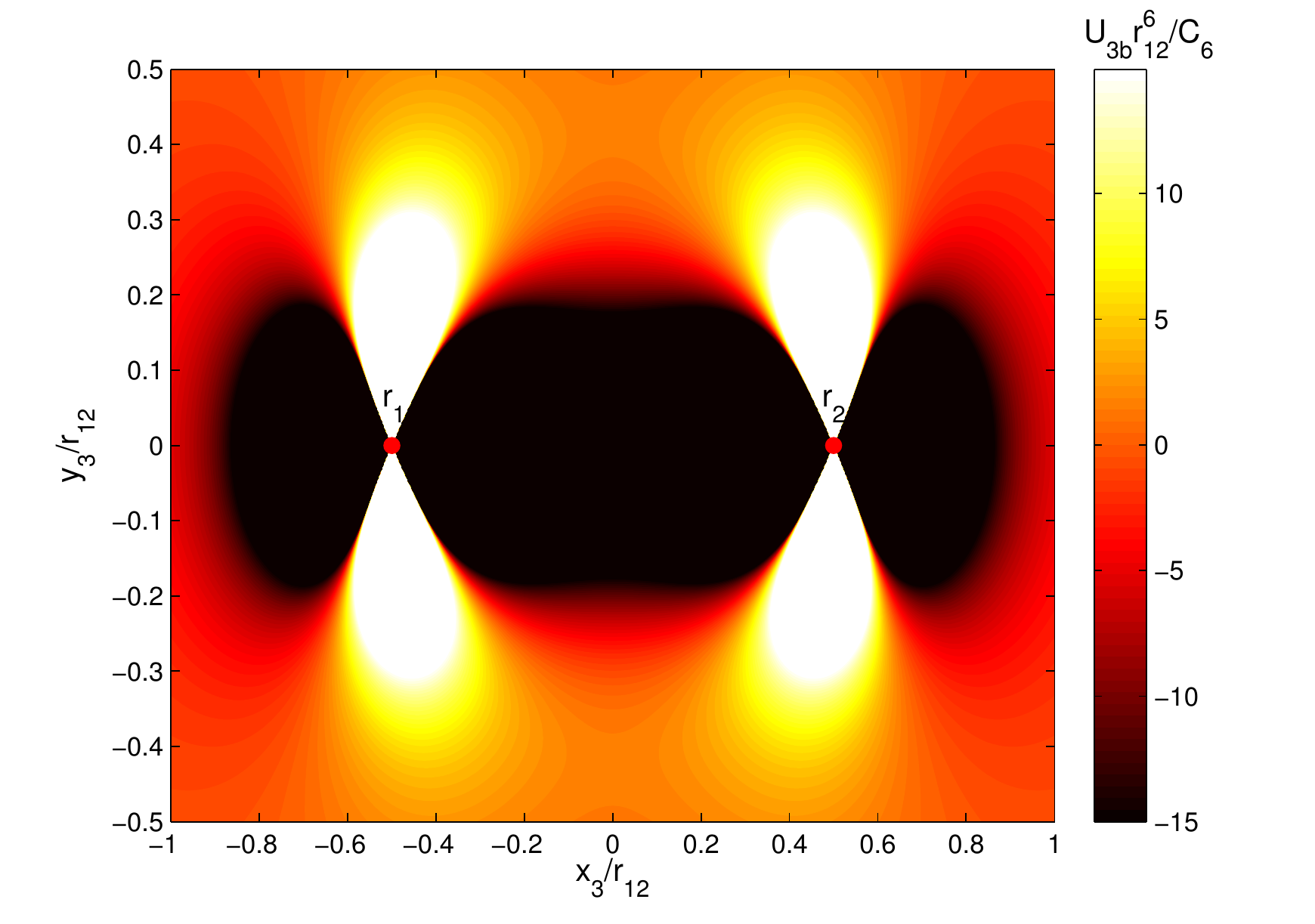}}
\caption{Three-body interaction scaled with $C_6/r_{12}^6$ for a three-body compound. The positions of two particles are 
fixed and are shown by red circles at $r_1$ and $r_2$ in the figure.}\label{fig7}
\end{center}
\end{figure}

In \fref{fig7}, we show the interaction potential scaled by the bare van der Waals potential $C_6/r_{12}^6$ for three particles at positions ${\bf r}_i$ ($i=1,2,3$), fixing ${\bf r}_1$ and ${\bf r}_2$ and varying the position ${\bf r}_3$ of the third atom. The interaction potential exhibits a dipolar pattern, independent of the orientation of the three particle compound. Such genuine three-body interactions have attracted great theoretical interest, as they give rise to a wealth of exotic phenomena in condensed matter systems \cite{moessner01, motrunich02}. While systems with dominating multi-body interactions are typically scarce in nature, lattices of ultracold polar molecules have been proposed for realizing such conditions \cite{buechler07p}. Recent experiments have demonstrated the implementation of elementary spin models via Rydberg excitation of atomic lattices \cite{betelli13}, such that the combination with the present microwave control scheme may offer a viable approach to artificial quantum magnets with dominating three-body interactions. 

\section{Conclusion}
In summary, we have shown that off resonant dressing of Rydberg states by multiple microwave fields provides a versatile approach to control Rydberg-Rydberg atom interactions. In addition to inducing familiar dipole-dipole interactions, the presented scheme enables to induce isotropic interactions and effective van der Waals potentials, allowing to turn off the long-range tail of the field-free interactions entirely, for proper parameters realizes conditions with dominating three-body interactions. For simplicity, we have neglected the fine structure splitting of the Rydberg states, which is well justified for describing the van der Waals interaction between ($ns,ns$) pairs \cite{singer05,walker08}. Including  fine structure for the microwave dressing one obtains identical microwave-induced potentials, upon adding two dressing fields with circular polarisation opposite to that of the $\sigma$- and $\sigma^\prime$-fields (cf. Fig.\ref{fig1}) in order to assure symmetric dressing for the $m_J=\pm1/2$ states of the $nS_{1/2}$ Rydberg manifold.

Such control capabilities may find direct applications in ongoing experiments towards quantum simulations of magnetism using ultracold Rydberg ensembles. For example, the demonstrated reduction of the Rydberg blockade radius suggests a promising approach to significantly increase the number of excitable atoms to high lying, and thus long lived states in finite atomic lattices, as studied in recent experiments. The ability, to tune the long-distances behaviour of atomic interactions may also be important for experimental studies of nonequilibrium phase transitions in driven, dissipative Rydberg gases and lattices \cite{schauss12} for which the power-law tail of the van der Waals potential where recently shown to play an important role \cite{zoubi14}.  

Precise measurements of van der Waals Rydberg-Rydberg atom interactions have been recently demonstrated in experiments with two or three atoms confined in separate dipole traps at well controllable distances \cite{beguin13}. Such settings appear to be ideally suited to explore the effects of microwave dressing, optimize the control parameters and study the emergence of exotic genuine multi-body interactions due to microwave dressing.

Another important application of Rydberg-state interactions is the realization of large optical nonlinearities in cold atomic gases under electromagnetically induced transparency (EIT) conditions, which has been explored in recent theoretical and experimental studies \cite{maxwell13,parades14,sev11,petrosyan11,moha08, pritchard10,ates11}. It is proposed that interactions between atoms in the Rydberg states diminish the EIT effect and thereby leads to highly nonlinear absorption and refraction, corresponding to dissipative and unitary effective photon interactions\cite{sev11,petrosyan11,gorshkov11,parigi12,gorshkov13} . 
 Microwave control of the atomic interactions can therefore be employed to control photonic interactions  in such Rydberg-EIT media as shown in a recent experiment \cite{maxwell14}. 
\ack
We acknowledge the financial support from the EU through the FET-Open Grant MALICIA No. 265522, through the FET grant HAIRS No. 612862 and through the Marie Curie ITN "COHERENCE", and from IARPA through the project grant AQuA-64.\\ 

\appendix

\section{Rydberg-Rydberg atom interaction in rotating frame}\label{apa}
The Rydberg-Rydberg atom interaction Hamiltonian in the rotating frame of reference can be conveniently written as a sum of two terms
\begin{equation}
\hat{H}_{\rm AA}=\sum_{i<j}\hat{A}_{ij}+\hat{B}_{ij}
\end{equation}
where $\hat{A}_{ij}=\hat{A}_{x_ix_j}+\hat{A}_{y_iy_j}+\hat{A}_{z_iz_j}+\hat{A}^{\prime }_{x_ix_j}+\hat{A}^{\prime }_{y_iy_j}+\hat{A}^{\prime }_{z_iz_j}$, $\hat{B}_{ij}=\hat{B}_{x_iy_j}+\hat{B}_{x_iz_j}+\hat{B}_{y_iz_j}+\hat{B}^{\prime }_{x_iy_j}+\hat{B}^{\prime }_{x_iz_j}+\hat{B}^{\prime }_{y_iz_j}$ and 

\begin{eqnarray}
\fl\hat{A}_{x_ix_j}&=&\frac{\mu^2}{2R_{ij}^3}\left(1-3\frac{X_{ij}^2}{R_{ij}^2}\right)\left(\hat{\sigma}_{sp_-}^{(i)}\hat{\sigma}_{p_-s}^{(j)}-\hat{\sigma}_{sp_-}^{(i)}\hat{\sigma}_{p_+s}^{(j)}+\hat{\sigma}_{sp_+}^{(i)}\hat{\sigma}_{p_+s}^{(j)}-\hat{\sigma}_{sp_+}^{(i)}\hat{\sigma}_{p_-s}^{(j)}+{\rm h.c.}\right)\nonumber\\
\fl\hat{A}_{y_iy_j}&=&\frac{\mu^2}{2R_{ij}^3}\left(1-3\frac{Y_{ij}^2}{R_{ij}^2}\right)\left(\hat{\sigma}_{sp_-}^{(i)}\hat{\sigma}_{p_-s}^{(j)}+\hat{\sigma}_{sp_-}^{(i)}\hat{\sigma}_{p_+s}^{(j)}+\hat{\sigma}_{sp_+}^{(i)}\hat{\sigma}_{p_+s}^{(j)}+\hat{\sigma}_{sp_+}^{(i)}\hat{\sigma}_{p_-s}^{(j)}+{\rm h.c.}\right)\nonumber\\
\fl\hat{A}_{z_iz_j}&=&\frac{\mu^2}{R_{ij}^3}\left(1-3\frac{Z_{ij}^2}{R_{ij}^2}\right)\left(\hat{\sigma}_{sp_0}^{(i)}\hat{\sigma}_{p_0s}^{(j)}+{\rm h.c.}\right)\\
\fl\hat{B}_{x_iy_j}&=&3i\mu^2\frac{X_{ij}Y_{ij}}{R_{ij}^5}\left(\hat{\sigma}_{sp_+}^{(i)}\hat{\sigma}_{p_-s}^{(j)}-\hat{\sigma}_{sp_-}^{(i)}\hat{\sigma}_{p_+s}^{(j)}-{\rm h.c.}\right)\nonumber\\
\fl\hat{B}_{x_iz_j}&=&-3\mu^2\frac{X_{ij}Z_{ij}}{\sqrt{2}R_{ij}^5}\left(\hat{\sigma}_{sp_-}^{(i)}\hat{\sigma}_{p_0s}^{(j)}-\hat{\sigma}_{sp_+}^{(i)}\hat{\sigma}_{p_0s}^{(j)}+\hat{\sigma}_{sp_0}^{(i)}\hat{\sigma}_{p_-s}^{(j)}-\hat{\sigma}_{sp_0}^{(i)}\hat{\sigma}_{p_+s}^{(j)}+{\rm h.c.}\right)\nonumber\\
\fl\hat{B}_{y_iz_j}&=&3i\mu^2\frac{Y_{ij}Z_{ij}}{\sqrt{2}R_{ij}^5}\left(\hat{\sigma}_{sp_-}^{(i)}\hat{\sigma}_{p_0s}^{(j)}+\hat{\sigma}_{sp_+}^{(i)}\hat{\sigma}_{p_0s}^{(j)}-\hat{\sigma}_{sp_0}^{(i)}\hat{\sigma}_{p_-s}^{(j)}-\hat{\sigma}_{sp_0}^{(i)}\hat{\sigma}_{p_+s}^{(j)}-{\rm h.c.}\right)
\end{eqnarray}

The operators  $\hat{A}^{\prime }$ and $\hat{B}^{\prime }$ are obtained by replacing $\mu\rightarrow\mu^{\prime}$ and $p_{0,\pm}\rightarrow p_{0,\pm}^{\prime} $.  

\section{Three-Body Interaction terms}\label{apb}

Similarly we write the three-body interaction in (\ref{E4}) as $U_{\rm 3b}({\bf R}_i,{\bf R}_j,{\bf R}_k) = E_{ijk}+E_{kij}+E_{jki}$, where
\begin{equation}
 E_{ijk}=G_{ijk}-G^{\prime}_{ijk},
\end{equation}
and

\begin{eqnarray}
 \fl G_{ijk}&=&\frac{\mu^4}{\Delta R_{ij}^3R_{jk}^3}\left[\left(\frac{\Omega_\pi}{\Delta}\right)^2(\alpha_{ij}\alpha_{jk}+\gamma_{ij}\gamma_{jk}^*+\gamma_{ij}^*\gamma_{jk})+2\left(\frac{\Omega_\sigma}{\Delta}\right)^2(\epsilon_{ij}\epsilon_{jk}+\beta_{ij}\beta_{jk}^*+\gamma_{ij}^*\gamma_{jk})\right.\nonumber\\
\fl &+&\left.\sqrt{2}\frac{\Omega_\pi\Omega_\sigma}{\Delta^2}(\alpha_{ij}\gamma_{jk}+\alpha_{ij}\gamma_{jk}^*+\epsilon_{ij}\gamma_{jk}+\epsilon_{ij}\gamma_{jk}^*-\beta_{ij}\gamma_{jk}^*-\beta_{ij}^*\gamma_{jk})\right],
\end{eqnarray}

with
\begin{eqnarray}
 \alpha_{ij}&=&\left(1-\frac{3Z_{ij}^2}{R_{ij}^2}\right),\nonumber\\
 \beta_{ij}&=&\frac{3}{2R_{ij}^2}(X_{ij}^2-Y_{ij}^2-2iX_{ij}Y_{ij}),\nonumber\\
\gamma_{ij}&=&-\frac{3}{2R_{ij}^2}(X_{ij}Z_{ij}-iY_{ij}Z_{ij}),\nonumber\\
 \epsilon_{ij}&=&\left(1-\frac{3(X_{ij}^2+Y_{ij}^2)}{2R_{ij}^2}\right).
\end{eqnarray}
The contribution from the lower transition $G^{\prime}_{ijk}$ is obtained by replacing $\mu\rightarrow\mu^{\prime}$, $\Omega_\pi\rightarrow\Omega^{\prime}_\pi$ and  
$\Omega_\sigma\rightarrow\Omega^{\prime}_\sigma$. $E_{kij}$ and $E_{jki}$ are written in the same form.


\section*{References}

\begin{thebibliography}{10}

\bibitem{lfc01}  Lukin M D, Fleischhauer M, Cote R, Duan  L M, Jaksch D, Cirac J I and Zoller P 2001 {\it Phys. Rev. Lett. } {\bf 87} 037901
\bibitem{jaksch00} Jaksch D, Cirac J I, Zoller  P, Rolston S L, Cote R and Lukin M D 2000 {\it Phys. Rev. Lett.} {\bf 85} 2208
\bibitem{saffman02} Saffman M and Walker T G 2002 {\it Phys. Rev. A} {\bf 66} 065403 
\bibitem{dudin12} Dudin Y O and Kuzmich A 2012 {\it Science} {\bf{336}} 887 
\bibitem{weimer10}  Weimer H,  M\"{u}ller M, Lesanovsky I, Zoller P and B\"{u}chler H P  2010 {\it Nat. Phys.} {\bf 6} 382 
\bibitem{pode10} Pohl T, Demler E and Lukin M D 2010 {\it Phys. Rev. Lett.} {\bf 104} 043002
\bibitem{bouchoule02}  Bouchoule I and M\o{}lmer K 2002 {\it Phys. Rev. A} {\bf 65} 041803(R)
\bibitem{hofmann13}Hofmann C S, G\"{u}nter G, Schempp H, Robert-de-Saint-Vincent M, GŠrttner M, Evers J, Whitlock S and Weidem\"{u}ller M 2013 {\it Phys. Rev. Lett.} {\bf{110}} 203601
\bibitem{peyronel12} Peyronel T, Firstenberg O, Liang Q -Y, Hofferberth S, Gorshkov A V, Pohl T, Lukin M D and Vuleti\'{c} V 2012 {\it Nature} {\bf{488}} 57 
\bibitem{gallagher} Gallagher T F 1994 {\it Rydberg atoms} (Cambridge Univ. Press)
\bibitem{li05}  Li W, Tanner P J and Gallagher T F 2005 {\it Phys. Rev. Lett.} {\bf 94} 173001
\bibitem{afrousheh04}  Afrousheh K, Bohlouli-Zanjani P, Vagale D, Mugford A, Fedorov M and Martin J D D 2004 {\it Phys. Rev. Lett.} {\bf 93} 233001
\bibitem{mudrich05} Mudrich M, Zahzam N, Vogt T, Comparat D and Pillet P 2005 {\it Phys. Rev. Lett.} {\bf 95} 233002 
\bibitem{carroll04} Carroll T J, Claringbould K, Goodsell A, Lim M J and Noel M W 2004 {\it Phys. Rev. Lett.} {\bf 93} 153001
\bibitem{pss09} Pohl T, Sadeghpour H R, Schmelcher P 2009 {\it Phys. Rep.} {\bf 484} 181
\bibitem{bz07} Bohlouli-Zanjani P, Petrus J A and Martin J D D 2007 {\it Phys. Rev. Lett.} {\bf 98} 203005
\bibitem{petrus08} Petrus J A, Bohlouli-Zanjani P and Martin J D D 2008 {\it J. Phys. B} {\bf 41} 245001
\bibitem{han09} Han J and Gallagher T F 2009 {\it Phys. Rev. A} {\bf 79} 053409
\bibitem{park11} Park  H, Tanner P J, Claessens B J, Shuman E S and Gallagher T F 2011 {\it Phys. Rev. A} {\bf 84} 022704 
\bibitem{buechler07p} Buechler H P, Micheli A and Zoller P 2007 {\it Nat. Phys.} 2007 {\bf 3} 726
\bibitem{buechler07n} Buechler H P, Demler E, Lukin M, Micheli A, Prokof'ev N, Pupillo G and Zoller P 2007 {\it Phys. Rev. Lett.} {\bf 98}, 060404
\bibitem{gorshkov08} Gorshkov A V, Rabl P, Pupillo G, Micheli A, Zoller P, Lukin M D and B\"uchler H P,  2008 {\it Phys. Rev. Lett.} {\bf 101} ,073201
\bibitem{muller08} M\"{u}ller M, Liang L, Lesanovsky I and Zoller P 2008 {\it N. J. Phys.} {\bf 10} 093009
\bibitem{tanas11} Tanasittikosol M, Pritchard J D, Maxwell D, Gauguet A, Weatherill K J, Potvliege R M and Adams C S 2011 {\it J. Phys. B} {\bf 44} 184020
\bibitem{moha07}Mohapatra A K, Jackson T R and Adams C S 2007  {\it Phys. Rev. Lett.} {\bf{98}} 113003
\bibitem{bason08}Bason M G, Mohapatra A K, Weatherill KJ and Adams C S 2008 {\it Phys. Rev. A} {\bf 77} 032305
\bibitem{maxwell13}Maxwell D, Szwer D J, Paredes-Barato D, Busche H, Pritchard J D, Gauguet A, Weatherill K J,
Jones M P A and Adams C S 2013  {\it Phys. Rev. Lett.} {\bf 110} 103001
\bibitem{parades14} Paredes-Barato D and Adams C S 2014 {\it Phys. Rev. Lett.} {\bf 112} 040501
\bibitem{jones13} Jones L A, Carter J D and Martin J D D, 2013 {\it Phys. Rev. A} {\bf 87} 023423
\bibitem{colombo13}Colombo  A P, Zhou Y, Prozument K, Coy S L and Field R W 2013 {\it J. Chem. Phys.} {\bf 138} 014301
\bibitem{saffman10} Saffman M, Walker T G and M\o lmer K 2010 {\it Rev. Mod. Phys.} {\bf 82} 2313
\bibitem{comparat10}Comparat D and Pillet P 2010  {\it J. Opt. Soc. Am.} {\bf{27}} A208
\bibitem{zimmerman79}Zimmerman M L, Littman M G, Kash M M and Kleppner D 1979 {\it Phys. Rev. A} {\bf20} 2251
\bibitem{li03}Li W, Mourachko I, Noel M W and Gallagher T F 2003 {\it Phys. Rev. A} {\bf 67} 052502
\bibitem{chu04} Chu S-I and Telnov D A 2004 {\it Phys. Rep.} {\bf 390} 1
\bibitem{singer05} Singer K, Stanojevic J, Weidem\"uller M and Cote R 2005 {\it J. Phys. B} {\bf 38} S295
\bibitem{walker08} Walker T G and Saffman M 2008 {\it Phys. Rev. A} {\bf 77} 032723 
\bibitem{bois02}Boisseau C, Simbotin I and C\^{o}t\'{e} R 2002 {\it Phys. Rev. Lett.} {\bf{88}} 133004
\bibitem{cooper09}Cooper N R and Shlyapnikov G V 2009 {\it Phys. Rev. Lett.} {\bf 103} 155302 
\bibitem{gallmol}Kiffner M, Park H, Li W and Gallagher T F 2012 {\it Phys. Rev. A} {\bf86} 031401(R)
\bibitem{moessner01} Moessner R and Sondhi S L 2001 {\it Phys. Rev. Lett.} {\bf86} 1881
\bibitem{motrunich02}Motrunich O I and  Senthil T 2002 {\it Phys. Rev. Lett.} {\bf89} 277004 
\bibitem{betelli13}Bettelli S, Maxwell D, Fernholz T, Adams C S, Lesanovsky I and Ates C 2013 {\it Phys. Rev. A} {\bf88} 043436 
\bibitem{schauss12}Schauss P, Cheneau M, Endres M, Fukuhara T, Hild S, Omran A, Pohl T, Gross C, Kuhr S and Bloch I 2012 {\it Nature} {\bf491} 87
\bibitem{zoubi14}Zoubi H, Eisfeld A and W\"{u}ster S 2014 {\it Phys. Rev. A} {\bf89} 053426
\bibitem{beguin13}B\'{e}guin  L, Vernier A, Chicireanu R, Lahaye T and Browaeys A 2013 {\it Phys. Rev. Lett.} {\bf110} 263201
\bibitem{sev11} Sevin\c{c}li S, Henkel N, Ates C and Pohl T 2011 {\it Phys. Rev. Lett.} {\bf 107} 153001
\bibitem{petrosyan11}Petrosyan D, Otterbach J and Fleischhauer M 2011 {\it Phys. Rev. Lett.} {\bf107} 213601
\bibitem{moha08} Mohapatra A K, Bason M G, Butscher B, Weatherill K J and Adams C S 2008  {\it Nat. Phys.} {\bf 4} 890 
\bibitem{pritchard10} Pritchard J D, Maxwell D, Gauguet A, Weatherill K J, Jones M P A and Adams C S 2010 {\it Phys. Rev. Lett.} {\bf 105} 193603
\bibitem{ates11} Ates C, Sevin\c{c}li S and Pohl T 2011 {\it Phys. Rev. A} {\bf 83} 041802(R)
 \bibitem{gorshkov11} Gorshkov A V, Otterbach J, Fleischhauer M, Pohl T and Lukin M D 2011 {\it Phys. Rev. Lett.} {\bf107} 133602
\bibitem{parigi12}Parigi V, Bimbard E, Stanojevic J, Hilliard A J, Nogrette F, Tualle-Brouri R, Ourjoumtsev A and Grangier P 2012 {\it Phys. Rev. Lett.} {\bf109} 233602
\bibitem{gorshkov13} Gorshkov A V, Nath R and Pohl T 2013 {\it Phys. Rev. Lett.} {\bf110} 153601
\bibitem{maxwell14}Maxwell D, Szwer D J, Paredes-Barato D, Busche H, Pritchard J D, Gauguet A, Jones M P A and Adams C S 2014 {\it Phys. Rev. A} {\bf 89} 043827
\end{thebibliography}
\end{document}